\documentclass[preprint,12pt]{elsarticle}
\usepackage{amssymb}
\usepackage{graphicx}
\usepackage{epsfig}

\makeatletter
\def\ps@pprintTitle{%
 \let\@oddhead\@empty
 \let\@evenhead\@empty
 \def\@oddfoot{}%
 \let\@evenfoot\@oddfoot}
\makeatother

\begin{document}

\begin{frontmatter}

\title{Hierarchical organization of H. Eugene Stanley scientific collaboration community in weighted network representation}

\author[label1,label2]{Stanis{\l}aw Dro\.zd\.z}
\ead{stanislaw.drozdz@ifj.edu.pl}
\author[label1]{Andrzej Kulig}
\author[label1]{Jaros{\l}aw Kwapie\'n}
\author[label2]{Artur~Niewiarowski}
\author[label2]{Marek Stanuszek}
\address[label1]{Complex Systems Theory Department, Institute of Nuclear Physics, Polish Academy of Sciences, Krak\'ow, Poland}
\address[label2]{Faculty of Physics, Mathematics and Computer Science, Cracow University of Technology, Krak\'ow, Poland}

\begin{abstract}
By mapping the most advanced elements of the contemporary social interactions, the world scientific collaboration network develops an extremely involved and heterogeneous organization. Selected characteristics of this heterogeneity are studied here and identified by focusing on the scientific collaboration community of H.~Eugene Stanley - one of the most prolific world scholars at the present time. Based on the {\it Web of Science} records as of March 28, 2016, several variants of networks are constructed. It is found that the Stanley $\#1$ network - this in analogy to the Erd\H{o}s $\#$ - develops a largely consistent hierarchical organization and Stanley himself obeys rules of the same hierarchy. However, this is seen exclusively in the weighted network representation. When such a weighted network is evolving, an existing relevant model indicates that the spread of weight gets stimulation to the multiplicative bursts over the neighbouring nodes, which leads to a balanced growth of interconnections among them. While not exclusive to Stanley, such a behaviour is not a rule, however. Networks of other outstanding scholars studied here more often develop a star-like form and the central hubs constitute the outliers. This study is complemented by a spectral analysis of the normalised Laplacian matrices derived from the weighted variants of the corresponding networks and, among others, it points to the efficiency of such a procedure for identifying the component communities and relations among them in the complex weighted networks.
\end{abstract}


\end{frontmatter}

\section*{Highlights}

\begin{itemize}

\item Complexity characteristics of the scientific collaboration networks of several world renown scholars are studied.

\item Scientific collaboration community of H. Eugene Stanley self-organizes into a scale-free hierarchy but this is seen exclusively in the weighted network representation.

\item Such a network organization indicates that during its evolution the spread of weight gets stimulation to a balanced growth of interconnections among them.

\item Collaboration networks of other world's outstanding and prolific scholars often develop a star-like form and the central hubs constitute outliers.

\item Spectral decomposition of the normalised Laplacian matrices is shown to be efficient in disentangling internal community ties.

\end{itemize}

\section{Introduction}
\label{intro}

The accelerating process of world globalization embraces and pervades all aspects of the human activity. Contemporary means and standards of conducting the scientific investigations deserve a special attention in this context as their progress at the same time constitutes both the condition and the result of this world globalization process. Indeed, the world most advanced scientific contemporary initiatives are based on multinational and often even on highly multidisciplinary collaborations. Some of them, like the ones carrying out the high energy physics experiments at CERN and at DESY in Europe, at Fermilab and at Brookhaven in the US, at KEK in Japan or the ones conducting the global astronomical sky-observations, are largely administratively arranged as far as their organization and staff involved is concerned. Typically this predetermines the co-authorship composition, usually very numerous, of the resulting, also numerous, publications. However, there recently emerge more spontaneous and at the same time more dynamical forms of the scientific cooperation. In most cases they are driven by the contemporary interdisciplinary trends in research, such that they involve a group of renown scientists (or even a single one) who, by their ability to create a scientifically stimulating environment, attract others to a productive collaboration, which proliferates further through various disciplines and diversified co-authorship compositions (Adams, 2012).

Paul Erd\H{o}s, the famous Hungarian mathematician (De~Castro \& Grossman, 1999), who has written over 1400 papers with over 500 co-authors and who thus inspired the concept of the Erd\H{o}s number, can be considered a forerunner. At present an even more spectacular cascading of scientific collaboration of this kind can be observed. In this regard H.~Eugene Stanley, professor at the Boston University, whose scientific activity comprises a broad range of areas such as \textit{Aggregation, Viscous Fingering, Statistical Physics, Phase Transitions, Critical Phenomena, Granular Materials, Surface Physics, Econophysics, Chemistry, Water, Social Networks, Physiology, Medicine, and Neuroscience}, and his constantly increasing number of collaborators create a particularly interesting phenomenon to study. H.E.~Stanley's $h=125$ index due to ${\it N} = 1208$ published articles co-authored in total by 738 scientists, as on March 28, 2016, listed by the \textit{Web of Science (WoS)}, with all these figures constantly increasing (currently $h=134$ and ${\it N} = 1301$) provides a formal evidence of this great success and his scientific collaboration network (SCN) deserves thus a particular attention.

Studying characteristics of various aspects of the scientific collaboration potentially constitutes a significant contribution towards understanding the structure and dynamics of the social interactions (Luukkonen, Persson, \& Sivertsen, 1992; Katz, 1994; Grossman \& Ion, 1995; Jin, Girvan \& Newman, 2001; Liljeros et al., 2001; Jiang et al., 2013) but, first of all, it is of great importance for an efficient stimulation of the future science development (Wilsdon, 2011; Ausloos, 2013; Mi\'skiewicz, 2013; Bourgrine 2014; Ausloos, 2014a; Rotundo, 2014). Quantifying properties of the scientific collaboration networks in an informative and transparent way becomes highly facilitated (Barabasi et al., 2002; Li et al., 2007; Palla, Barab\'asi \& Vicsek, 2007; Lee, Goh, Kahng \& Kim, 2010; Liu, Xu, Small \& Chi, 2011) thanks to the great advances in the field of network theory (Albert \& Barab\'asi, 2002). Most of the existing related works study the global properties of the collaboration networks (de~Solla~Price, 1965; Wagner \& Leydesdorff, 2005; Wuchty, Jones \& Uzzi, 2007; Freeman, Ganguli \& Murciano-Goroff, 2014), including their evolutionary aspects (Newman, Strogatz \& Watts, 2001a; Newman, 2001b; Newman, 2001c; Newman, 2004; Tomassini \& Luthi, 2007), or occasionally point to the individual country contribution (He, 2009; Perc, 2010). Fewer works focus on characteristics of the selected scientists (Ding, 2011) in their creative role and of the range of their influence in the collaboration network. In order to make this issue and the related characteristics even more exposed, here, for several most outstanding scholar figures working in the domain of exact sciences, with a particular focus on H. Eugene Stanley, we generate their collaboration networks based exclusively on all the publications involving that particular scholar. Nodes then represent all the authors who appeared in any of the common publications and the links among them are assigned when their names appear together in the same publication. By construction, a node representing the author X defining such a network constitutes the central hub and all the other nodes in such a network have the collaboration number 1 relative to X, which by analogy to the Erd\H{o}s number can be termed the X number 1 (X $\#1$).

\section{Network construction and description}
\label{res}

All the results presented in this work have been obtained using the data downloaded from the \textit{Web of Science}. This website provides one of the most reliable and complete scientometrics sources. It covers many scientific disciplines belonging both to the exact sciences, to engineering as well as to the life sciences. Still, ensuring that all scientists are clearly identifiable and distinguishable, as needed in the present analysis, appears a highly non-trivial task. There are several elements that demand a special care. One particularly important is a proper distinction of different scientists. As it has already been estimated (Newman, Strogatz \& Watts, 2001a; Newman, 2001b; Perc, 2010), about $5\%$ of all scientists have the same initials and surnames. What is even more troublesome is that there exist different scientists having the same name and the same surname as well. In order to overcome such an equivocation, an additional criterion of the scientific affiliation has been applied. This of course helps, but does not resolve the problem entirely due to the significant mobility of scientists. Another problem is the presence of typos in the names and surnames. Such possible errors have been taken care of by using the \textit{Levenshtein measure} (Levenshtein, 1966) to strings of letters, here representing the names and surnames.

As in essentially all the network cases, the topology of SCN can be expressed by its adjacency matrix $\bf A$ whose elements $a_{ij}$ assume the value 1, thus express existence of the resulting link, if the authors $i$ and $j$ co-author at least one publication. Otherwise $a_{ij}$ equals 0. The corresponding $i$-th node degree $k_i = \sum_{j=1}^N a_{ij}$, where $N$ is the total number of authors (nodes) within the network. Complete description of SCN requires, however, taking into account not only its topology but also the weights of the links among the nodes (Newman, 2001c; Boccaletti et al., 2006). In SCN the weight of a given link is determined by the number $n_{ij}$ of publications co-authored by the $i$-th and $j$-th authors. The so-weighted $i$-th node degree, denoted as $k^w$, can be written as $k^w_i=\sum_{j=1}^N a_{ij}n_{ij}$. 

A more sophisticated way of introducing strengths of the collaborative ties is to account for the varying number $m_l$ of the co-authors of the corresponding publication $l$ by defining 
\begin{equation}
{\it s}_{ij}=\sum_l {{\delta^l_i\delta^l_j}\over{m_l-1}}, 
\label{strength} 
\end{equation}
where $\delta^l_i$ is 1 if the author $i$ is a co-author of the publication $l$ and zero otherwise and $l$ runs over all the publications involved (Newman, 2001c). Thus,
\begin{equation}
s_i=\sum_{j(\neq i)} {\it s}_{ij} 
\end{equation}
expresses the collaborative strength of the author $i$ since, as it can be easily verified by substitution, $s_i=\sum_k \delta^k_i$ just equals the number of papers that $i$ has co-authored with others. 
Distributions of the above three variants of the node degrees will be studied below. 

Another topologically informative network measure is the clustering coefficient, which for a node $i$ with $k_{i}$ links (edges) is defined as
\begin{equation}
C_{i}=2q_{i}/k_{i}(k_{i}-1),
\end{equation}
where $q_{i}$ is the number of edges between the $k_{i}$ neighbours of $i$.
In the case of a hierarchical network, the clustering coefficient of a node with $k$ links follows the scaling law $C(k) \sim k^{-1}$ (Ravasz \& Barab\'asi, 2003). Extension of the clustering coefficient concept to incorporate weights is not unique, however. In the literature there exist several equally acceptable definitions, but all among themselves they result in the different distributions and therefore will not be considered here. 

\section{Stanley's scientific collaboration network}
\label{HES}

A sketch of the network central to the present study is shown in Fig.~1. This is the H. Eugene Stanley's (HES) scientific collaboration network where 738 nodes represent the scientists co-authoring publications with HES up to March 28, 2016. By analogy to the Erd\H{o}s number, these are thus  all scientists whose Stanley number equals 1. Links are drawn between the nodes if there exists a publication co-authored by the corresponding scientists. By construction the node representing HES is linked with all the other nodes and thus constitutes a central hub in this network. There are also direct links between other nodes and this reflects the presence of multiple-author publications. As some of the scientists co-authored many publications with HES in various author compositions, they give rise to several sub-hubs whose initials are explicitly indicated in Fig.~1. The numbers of common publications with HES (denoted as $L_{\rm HES}$), together with their full names, are listed in the Table~1.  
This Table lists also some other selected names from the Stanley $\#1$ network, their number of publications $L_{\rm HES}$ co-authored by HES, their entire number $L_{\rm TOT}$ of publications and, for those whose networks are explicitly drawn in Figs.~4-6, the numbers (in parentheses) of publications with no co-authors ($L_1$).

Explicitly indicated are also those lower-rank nodes in this particular network that represent other renown scientists whose own scientific collaboration networks have an interesting organization. Depending on the number of common publications, the links have different weights. This is taken into account in the lines thickness in Fig.~1.

\begin{figure}[!h]
\centering
\includegraphics[scale=0.3]{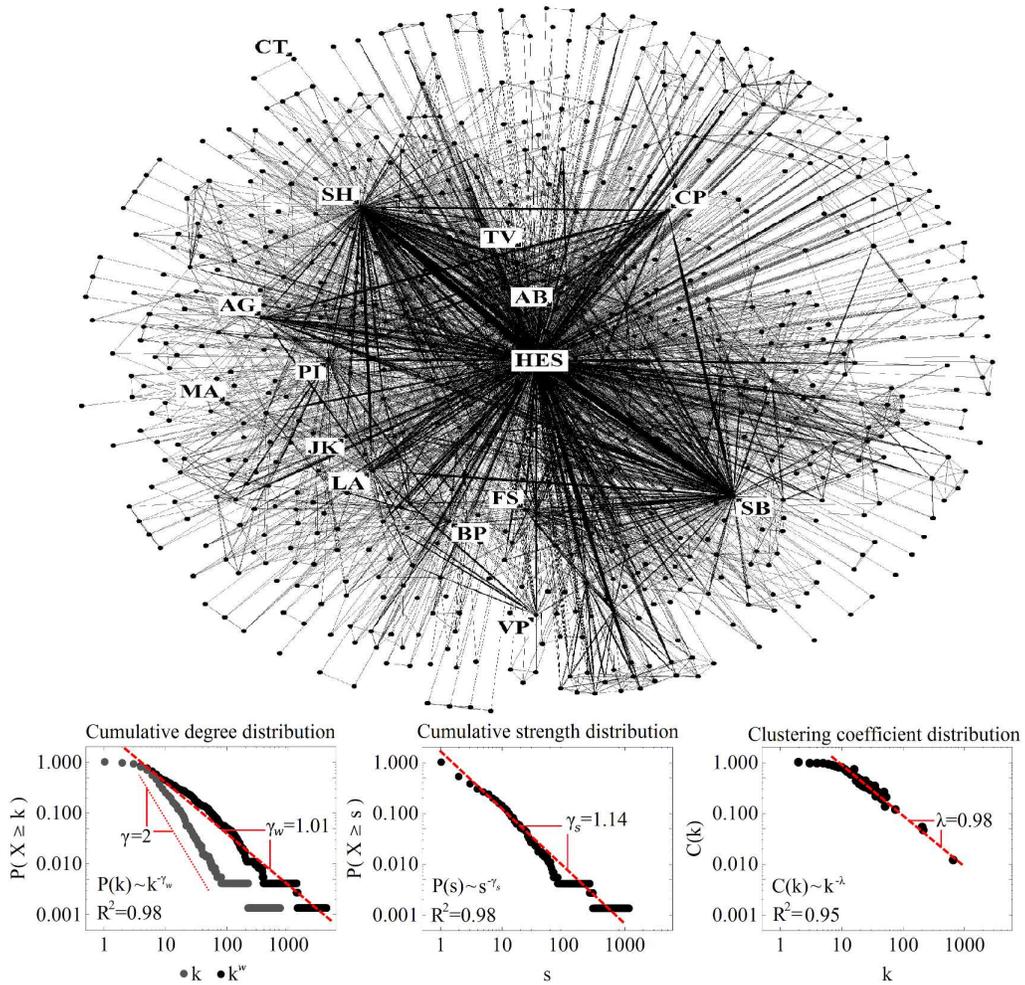}
\caption{H.~Eugene Stanley's (HES) scientific collaboration network (Stanley $\#1$) determined by all his 1208 publications listed by the {\it Web of Science} as of March 28, 2016. Nodes denote all co-authors of these publications and links are drawn between those authors  whose names appear in the same publication. The three lower panels include (i) the cumulative degree distributions, both unweighted $P(k)$ (gray dots) and weighted $P(k^w)$ (black dots), (ii) cumulative strength distributions $P(s)$ and (iii) the clustering coefficient $C(k)$ distribution, all characteristics for this network. Fits are indicated by the dashed lines, while the slope indicated by the dotted line serves guiding the eye.}
\label{fig1}
\end{figure}

The structure of the network in Fig.~1 already visually indicates its hierarchical organization. This organization appears, however, subtle, what finds quantitative evidence in terms of the relations between the three node-degree-related measures introduced above, i.e., $k$, $k^w$ and $s$. Their cumulative distributions, defined as  
\begin{equation}
P(X \geq x) \equiv \int_{x}^{\infty} P(x')dx' ,
\end{equation}
where $x$ denotes either $k$, $k^w$ or $s$, are shown in the log-log scale in the first and the second lower panels of this Fig.~1. The scaling exponents are evaluated by linear regression in the log-log scale using the \textit{Maximum Likelihood Estimation (MLE)} and the $R^2$ coefficient, as an error of this regression, reflects the goodness-of-fit statistics (Clauset, Shalizi \& Newman, 2009).

\begin{table}[!h]
\begin{center}
{\small
    \begin{tabular}{ | c | c |}
    \hline
    Author & $L_{\rm HES}/L_{\rm TOT} \ (L_1)$ \\
    \hline
    \hline
    Stanley, H.E. (HES) & 1208/1208 (47) \\
    \hline
    Havlin, S. (SH) & 307/697 (9) \\
    \hline
    Buldyrev, S.V. (SB) & 267/319 (4) \\
    \hline
    Amaral, L.A.N. (LA) & 81/165 \\
    \hline
    Sciortino, F. (FS) & 69/153 \\
    \hline
    Ivanov, P.C. (PI) & 65/117 \\
    \hline
    Peng, C.-K. (CP) & 62/158 \\
    \hline
    Goldberger, A.L. (AG) & 62/376 \\
    \hline
    Plerou, V. (VP) & 55/59 \\
    \hline
    Podobnik, B. (BP) & 49/118 \\
    \hline
    Barab\'asi, A.-L. (AB) & 22/263 (26) \\
    \hline
    Kert\'esz, J. (JK) & 11/243 \\
    \hline
    Vicsek, T. (TV) & 6/235 (25) \\
    \hline
    Ausloos, M. (MA) & 2/555 (41) \\
    \hline
    Tsallis, C. (CT) & 1/367 (60) \\
    \hline
    \end{tabular}
}
\end{center}
\label{tab1}
\caption{Statistics of H.E. Stanley (HES) and some of his collaborators: the number of all publications of a given author ($L_{\rm TOT}$), the number of publications coauthored by HES ($L_{\rm HES}$), and the number of considered monographs ($L_1$).}
\end{table}

Clearly, in all the three distributions there are segments where the straight line fits can be applied, pointing to some underlying scale-free effects that this network develops. At the same time, there are  several essential quantitative differences in the corresponding characteristics, however, especially between the unweighted ($P(k)$) and the weighted ($P(k^w)$, $P(s)$) cases. For $P(k)$ a straight line fit $P(X \geq k) \sim k^{-\gamma}$ applies in the $k$-interval of about $10 - 100$ with $\gamma \approx 2$ that in differential representation corresponds to 3.
This thus indicates that the intermediate $k$-degree nodes develop links that not only make them belonging to the preferential attachment universality class of networks, but even correspond exactly to the Barab\'asi-Albert model (Albert \& Barab\'asi, 2002). The central hub, HES, possesses, however, disproportionately more $k$-degrees and forms a clear outlier. Very interestingly, this effect disappears almost completely when the link weights, either expressed through $k_w$ or $s$, are taken into account. Within these two measures, the node degree distributions, including HES, tend to align along the same straight line. The corresponding best fit for $P(k^w)$ results in $\gamma_w \approx 1.01$ and for $P(s)$ in $\gamma_s \approx 1.14$. In fact, in $P(s)$ such a fit applies to all the scales and even the initial flattening of the Zipf-Mandelbrot type for the low-degree nodes seen in $P(k)$ and in $P(k^w)$, present also in other scientometrics analyses (Ausloos, 2014b), disappears. This resembles observations made in the linguistic context (Kulig, Kwapie\'n, Stanisz \& Dro\.zd\.z, 2017). There, by including the punctuation marks in the Zipfian analysis, in addition to words, corrects an analogous flattening such that the  Mandelbrot's amendment appears largely redundant and thus our language emerges as a more consistent composition on all the scales. The present result may thus be taken as an additional indication that the strength of the collaborative ties, as defined by  Eq.(1), offers the most consistent way of weighting the author's contributions.   

Overall hierarchical organization of the HES network is also confirmed by the coefficient $C(k)$ of a node with $k$ degrees whose scaling law $C(k) \sim k^{-1}$ (Ravasz \& Barab\'asi, 2003) asymptotically appears to be convincingly obeyed as it can be seen in the lower rightmost panel of Fig.~1.

\section{Networks of Erd\H{o}s and Witten}
\label{NEW}

In order to confront organization of the HES network with other possible organizations in this kind of networks, in Fig.~2 we show two analogous scientific collaboration networks: one for Paul Erd\H{o}s and another for the influential mathematical physicist Edward Witten, whose $h$-index of 131 in March 2016 (currently $h=134$) can be identified as the highest among active researchers representing or originating from the exact sciences. The scientometrics characteristics, including the numbers of articles published, the total numbers of co-authors in these published articles and the corresponding $h$-indices of these two scientists and of HES, correspondingly, are listed in Table~2.

\begin{center}
    \begin{tabular}{ | c | c | c | c | p{2cm} |}
    \hline
    * Based on \textit{Web of Science} & P.~Erd\H{o}s & E.~Witten &  H.E.~Stanley \\
    \hline
    \hline
    Number of articles & 1246 & 319 & 1208 \\
    \hline
    Number of collaborators & 391 & 140 & 738 \\
    \hline
    \textit{h}--index & 60 & 131 & 125 \\
    \hline
    \end{tabular}
\label{tab1}
\end{center}

The {\it Web of Science (WoS)} (http://www.webofknowledge.com) does not list all commonly recognized Erd\H{o}s' publications. A much more extended list of Erd\H{o}s' works, including all those listed by {\it WoS}, is provided by the Erd\H{o}s Number Project, which studies research collaborations among mathematicians and is maintained at the Oakland University (http://oakland.edu/enp/). Exceptionally, it is thus  this list, instead of {\it WoS}, which is used here to construct the Erd\H{o}s collaboration network. This source qualifies 1246 Erd\H{o}s' works as scientific publications, with 391 co-authors. At the same time this site notes that the total number of Erd\H{o}s publications equals 1525 with 511 co-authors; apparently not all fulfilling the criteria imposed. As far as Witten is concerned, all his recognized publications are listed by {\it WoS} including even three multi-author conference contributions.

\begin{figure}[!h]
\centering
\includegraphics[scale=0.55]{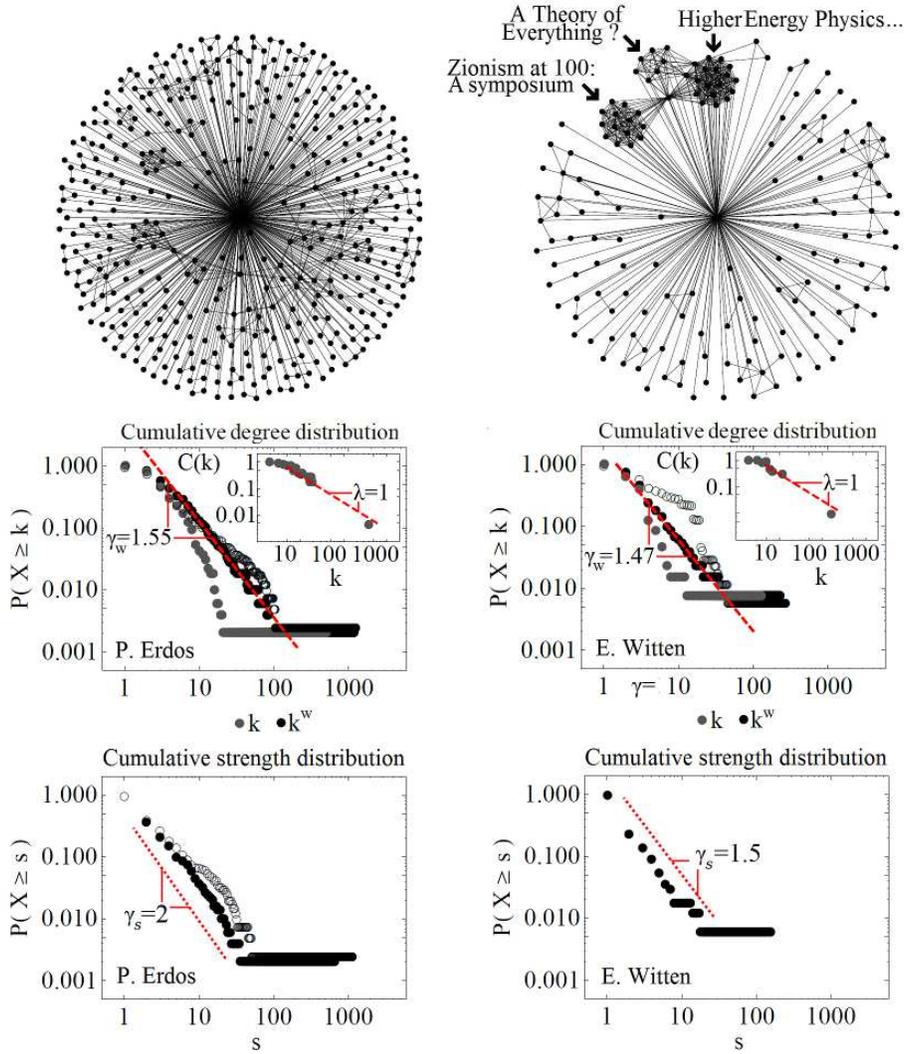}
\caption{Paul Erd\H{o}s' (left) and Edward Witten's (right) scientific collaboration networks, thus Erd\H{o}s $\#1$ and Witten $\#1$. Open dots in Witten's case indicate cumulative degree distribution $P(k)$ when the three clusters of nodes seen in this network are included. Otherwise the same convention as in Fig.~1 is used.}
\label{fig2}
\end{figure}

Clearly, topology of the two networks in Fig.~2 is not as extended in terms of the number of internal links as the one of HES. They both have a visibly dominant star-like component and scaling of the pure topology related $P(k)$ (gray dots in Fig.~2) can hardly be claimed. When the weights are taken into account, some partial scaling appears, however. In the Erd\H{o}s' network the weighted degree distribution $P(k^w)$ resulting from links among the other nodes displays an approximate scaling over almost two decades in $k^w$, with $\gamma_w \approx 1.55$. It has thus a lower intensity of weighted links than in the HES case. As a result the Erd\H{o}s' node still constitutes an outlier whose $k^w$ is by another one order of magnitude separated from all the others. Similarly, the clustering coefficient $C(k)$ is much compressed towards smaller $k$ as compared to HES and only one point representing Erd\H{o}s himself remains an outlier. Interestingly, however, it is placed not much below the slope of $\gamma=1$.

The Witten's node in his collaboration network is also an outlier, but essentially the links among other nodes give rise to any scaling neither in their weighted degree distribution $P(k^w)$ (open circles) nor in the clustering coefficient $C(k)$. This non-homogeneity in the Witten's network appears to be caused predominantly by the three multi-author publications mentioned above. Removing these three publications results in the distributions sketched by the black dots. Then the two characteristics become qualitatively similar to those of Erd\H{o}s with an even lower intensity of links as expressed by the corresponding $\gamma_w=1.47$ in the distribution of weighted degrees. As far as the range and quality of scaling is concerned, similar tendency is displayed by the strength $P(s)$ distribution shown in the lower panels of Fig.~2. It is thus evident that, contrary to the HES network, the two hubs, Erd\H{o}s' and Witten's, in their SCNs stay by far out of the distributions resulting from the node degrees of the other members of the corresponding networks.

For completeness one may here mention the two obvious extremes of the collaboration networks. On the one side, there is a case of no co-authors at all through the entire scientific activity (like, for instance, the Paul A.M. Dirac's one). The collaboration network then reduces itself to a trivial single node. On the opposite side, there are large collaborations of nearly fixed number of participants publishing papers always in the same author compositions. In the corresponding collaboration network any node is connected to all the remaining nodes and thus all of them have the same number of connections, thus the same degree. This kind of a network approximately represents large, predominantly administratively established collaborations.

\begin{figure}[!h]
\centering
\includegraphics[scale=0.45]{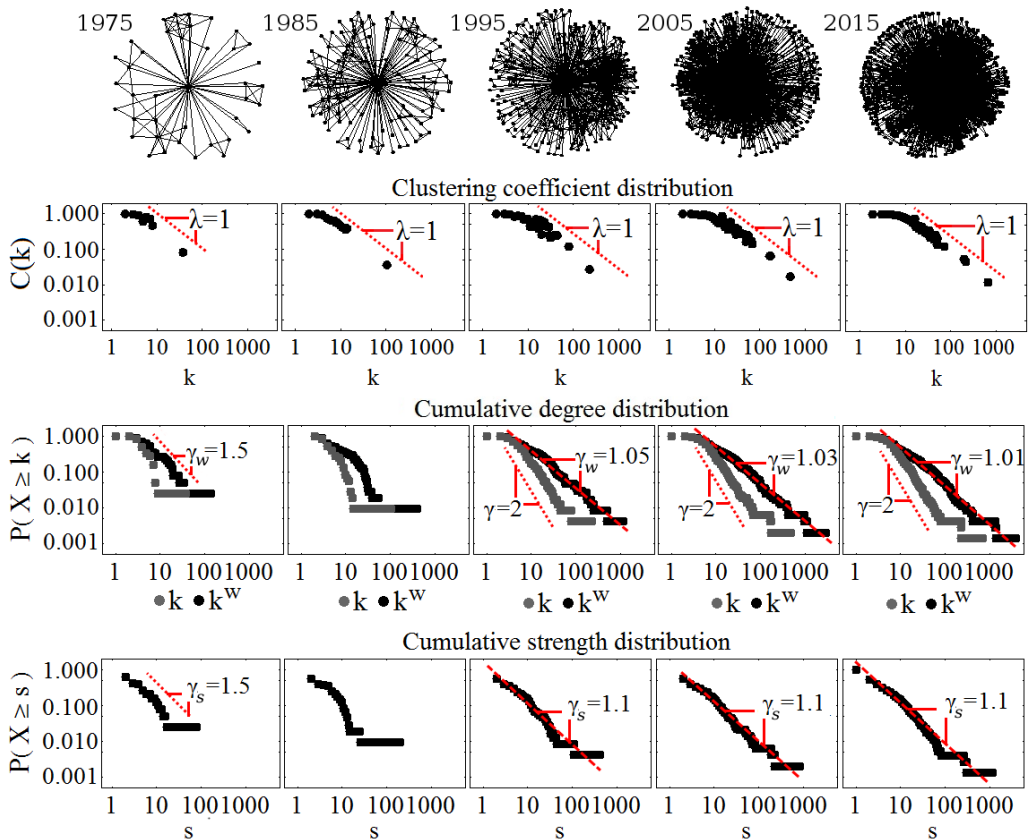}
\caption{Time evolution of H.~Eugene Stanley's (HES) scientific collaboration network (Stanley $\#1$) between 1975 and 2015 with the snapshots taken every 10 years, with the corresponding clustering coefficient and degree, both unweighted and weighted, distributions in the lower panels. The same convention as in Fig.~1 is used.}
\label{fig3}
\end{figure}

All the above possibilities indicate that the HES network, with its hierarchical organization through all levels, is an exception rather than a rule and that it therefore deserves a special attention, indeed. Such a collaboration network is of course a dynamical phenomenon and, definitely, some time is needed to attain not only this kind of richness but also a proper balance of links between all the participating nodes. The way the HES network has been evolving since the beginning of his scientific activity is illustrated in Fig.~3 with the five snapshots taken every 10 years between 1975 and 2015 with the corresponding degree, both unweighted and weighted, and the clustering coefficients distributions in the lower panels. Judging from these degree distributions a fully developed hierarchical organization had already been attained in around 1995.

\section{Networks of Ausloos, Barab\'asi, Buldyrev, Havlin, Tsallis, and Vicsek}

As it can be seen in Fig.~1, the two dominant sub-hubs in the HES network are those representing Shlomo Havlin (SH) and Sergey V. Buldyrev (SB), the long-term principal collaborators of H. Eugene Stanley. Their own collaboration networks, respectively Havlin $\#1$ and Buldyrev $\#1$, are thus sizeably overlapping with the one of HES. It is therefore natural to expect that they share some characteristics of its hierarchical organization. These two networks are shown in Fig.~4 together with the corresponding degree, both $k$ and $k^w$, strength $s$, and the clustering coefficient $C(k)$ distributions. Some parallels can easily be seen like, for instance, the fact that in both networks their central hubs, SH and SB, essentially align with the overall trend of the weighted degree distributions in these networks and, thus, they are not outliers. The quality of scaling of the weighted degree distributions is somewhat poorer, especially for SB, as compared to HES, but if one insists on fitting a single straight line, then the result for both $P(k^w)$ and $P(s)$ is consistent with $\gamma \approx 1$ in both networks, similarly as for HES. 

\begin{figure}[!h]
\centering
\includegraphics[scale=0.55]{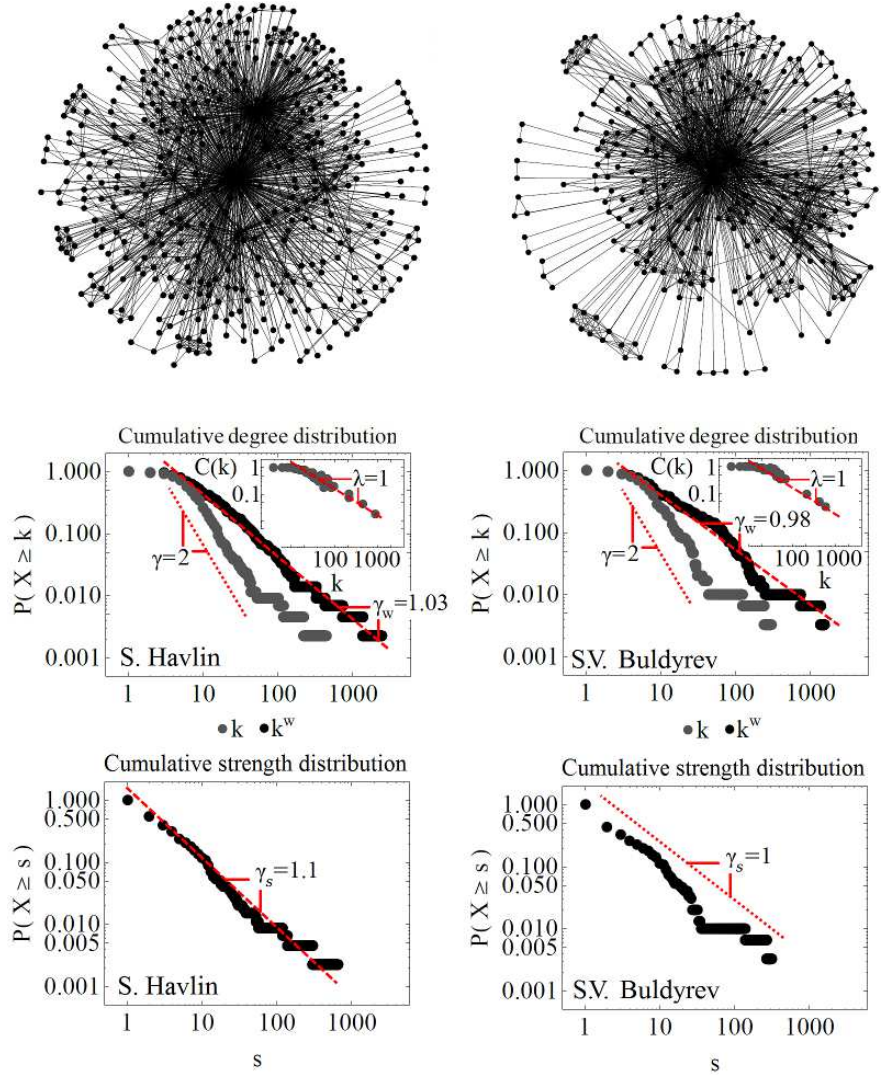}
\caption{Shlomo Havlin's (SH, left) and  Sergei V.~Buldyrev's (SB, right) scientific collaboration networks, thus Havlin $\#1$ and Buldrev $\#1$, with the corresponding cumulative unweighted and weighted degree distributions and the clustering coefficient distributions. The same convention as in Fig.~1 is used.}
\label{fig4}
\end{figure}

Among the nodes that appear in the HES scientific collaboration network, one can identify many extremely renown scholars. Several of them are explicitly indicated in Fig.~1 and among those some constitute further significant hubs in addition to the two shown in Fig.~4. The other nodes, rather peripheral in this network, are explicitly indicated for the reason that their own scientific collaboration networks also display a diverse organization. Four such cases, including Marcel Ausloos (MA), Albert-L\'aszl\'o Barab\'asi (AB), Constantino Tsallis (CT), and Tam\'as Vicsek (TV), are shown in Figs.~5 and 6. 

Clearly, out of these four, the most homogeneous hierarchical organization is revealed by the scientific collaboration network of Marcel Ausloos (Ausloos $\#1$), whose weighted degree distributions, in both $k^w$ and $s$, scale over more than two decades with the scaling exponents, $\gamma_w$ and $\gamma_s$, very close to unity. In this respect, it resembles most the HES case over the period 1995-2005. The three dominant sub-hubs in the MA network are due to Rudi Cloots (143 common publications with MA), Nicolas Vandewalle (75 common publications), Phillipe Vanderbemden (62 common publications), and Andr\'e Rulmont (54 common publications). Similarly, the clustering coefficient of a node with $k$ links for the Ausloos network follows the scaling law $C(k) \sim k^{-1}$ with quality comparable to the HES case. 

The Barab\'asi's network (Barab\'asi $\#1$), on the other hand, develops no homogeneous hierarchical organization as both the lack of scaling in all three variants of the node degree and in the clustering coefficient distributions indicate. This, actually, can be inferred directly from the structure of this network as its concentrations of nodes originate from the sparsely connected clusters, some of them mediated exclusively by the central hub, here Barab\'asi himself. 

\begin{figure}[!h]
\centering
\includegraphics[scale=0.55]{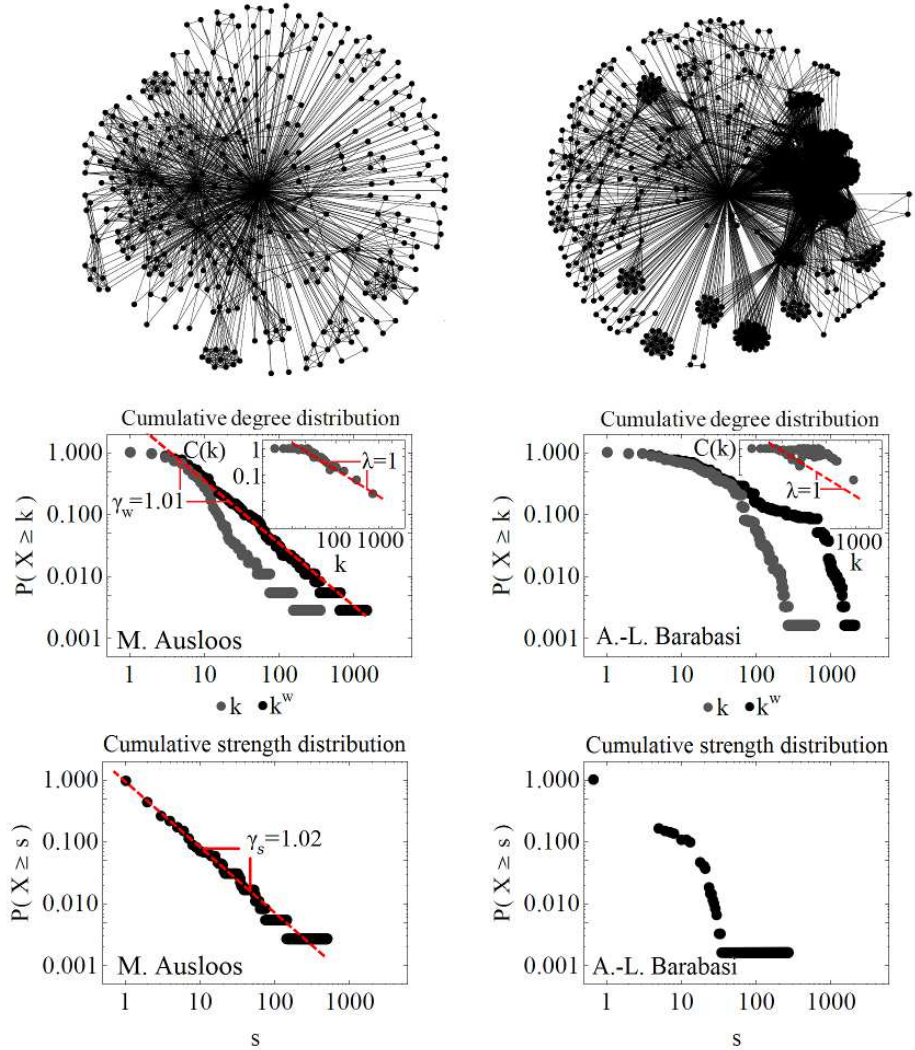}
\caption{Marcel Ausloos' (MA, left) and  Albert-L\'aszl\'o Barab\'asi's (AB, right) scientific collaboration networks, thus Ausloos $\#1$ and Barab\'asi $\#1$, with the corresponding cumulative unweighted and weighted degree distributions and the clustering coefficient distributions. The same convention as in Fig.~1 is used.}
\label{fig5}
\end{figure}

The other two networks, Tsallis $\#1$ and Viscek $\#1$, qualitatively resemble the Erd\H{o}s' star-like network with a sparser connectivity of the nodes in the Tsallis' network than in the Vicsek's one as schematically (since scaling is approximate here) indicated by the straight lines with the slopes $\gamma_w \approx 1.4$ (Tsallis') and $\gamma_w \approx 1.3$ (Vicsek's). The central hubs constitute outliers separated from all other nodes in both cases. Consistently, the clustering coefficients develop similar distributions as in the Erd\H{o}s' case.

\begin{figure}[!h]
\centering
\includegraphics[scale=0.6]{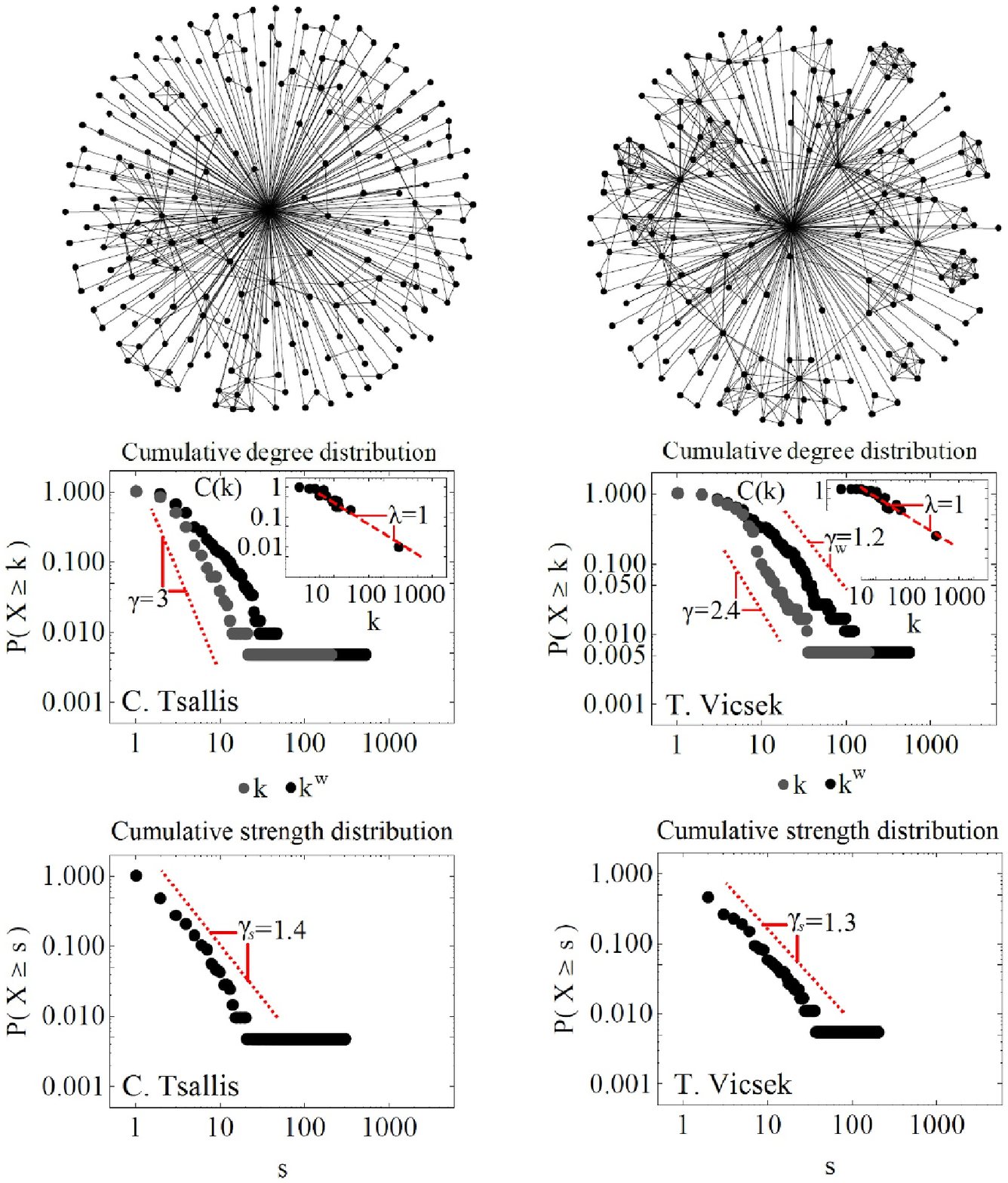}
\caption{Constantino Tsallis' (CT, left) and Tam\'as Vicsek's (TV, right) scientific collaboration networks, thus Tsallis $\#1$ and Vicsek $\#1$, with the corresponding cumulative unweighted and weighted degree distributions and the clustering coefficient distributions. The same convention as in Fig.~1 is used.}
\label{fig6}
\end{figure}

\section{Relation to weighted network model}

Comparison of the above results clearly shows that it is the weighted network representation, which allows more complete and informative disentangling of local differences in the scientific collaboration networks and, thus, offers scientifically more advanced framework. Modelling the observed characteristics of such networks is, however, much more difficult as compared to the unweighted cases. The main reason for this difficulty is that their growth involves mutual influence of the dynamics of links and weights and also some possible elements of the accelerated growth (Dorogovtsev \& Mendes, 2002) allowing appearance of the new links between the already existing nodes. Apparently, it is for these reasons that no sufficient progress has been attained so far in the rigorous modeling of complex weighted networks.

The most closely related existing model of weight driven growing networks (Barrat, Berthelemy \& Vespignani, 2004), even though still definitely much simplified in relation to the dynamics underlying evolution of the networks considered here, offers some preliminary unified view and quantifies ranges for the exponents of the degree distributions. In this model the usual preferential attachment (Barab\'asi \& Albert, 1999) is extended to the rule "busy get busier" (Barthelemy, Barrat, Pastor-Satorras \& Vespignani, 2005), in which new nodes connect more likely to the nodes carrying larger weights and being more central in terms of the interaction strength. Accordingly, the local rearrangements of weights between $i$ and its neighbours $j$ obeys the simple rule
\begin{equation}
w_{ij} \rightarrow w_{ij} + \Delta w_{ij}, \qquad \textrm{where} \qquad \Delta w_{ij} = \delta {w_{ij} \over s_i},
\label{delta}
\end{equation}
which means that a new link with the node {\it i} induces a total increase of activity $\delta$ that is proportionally distributed among the links departing from this node. 

The model thus involves only one parameter $\delta$ that reflects the fraction of weight transmitted by the new link onto the others. $\delta < 1$ corresponds to such a situation that a new connection does not lead to a more intense activity on existing links. In particular, for $\delta = 0$ the arrival of a new link does not affect the existing weights and this model becomes topologically equivalent to the Barab\'asi-Albert model (Barab\'asi, 1999). On the other hand, for $\delta > 1$ a new link  multiplicatively bursts the weight on neighbours. Within this model, in the large time limit, one obtains a power law scaling of the weighted degree distribution with the scaling exponent for the cumulative distribution
\begin{equation}
\gamma_s = 1 + {1 \over {2 \delta +1}},
\end{equation} 
with its values $\gamma_s \in [1,2]$. 

In the networks considered above, by their construction, all the new links involve the central hubs. From the perspective of this model, a value of the corresponding empirically determined $\gamma_s$ can be viewed as an indication of how such 'centers of condensation' stimulate mutual interactions of their nearest neighbours. HES with his $\gamma_s \approx 1.1$ and, thus, $\delta$ of about 5 appears very stimulative. Interestingly, the same applies to Marcel Ausloos though his (Ausloos $\#1$) network is not as large. However, it weekly overlaps with the HES network (unlike Havlin's and Buldyrev's) and can thus be considered independently. 

The opposite can be inferred for the networks from Figs. 2 and 6. Their central hubs with $\gamma_s < 1$ appear largely neutral in influencing interactions among the nearest neighbours. As far as this kind of neutrality is concerned, an extreme case is the Erd\H{o}s' SCN, whose $\delta$ consistent with $\gamma_s$ observed would be close to zero.

\section{Spectral decomposition of Laplacian matrices} 

The adjacency matrix $\bf A$ records all the information about nodes and how they are interconnected. The most mathematically consistent way of formulating this information is in terms of the Laplacian matrix (Bapat, 2014)
\begin{equation}
{\bf L} = {\bf D} - {\bf A},
\label{laplace}
\end{equation}
where $\bf D$ is a diagonal matrix composed of the nodes' weighted degrees. Equivalently, a normalized Laplacian matrix
\begin{equation}
{\bf {\hat L}} = {\bf D}^{-1/2} {\bf L} {\bf D}^{-1/2} = {\bf I} - {\bf D}^{-1/2} {\bf A} {\bf D}^{-1/2}
\label{nlaplace}
\end{equation}
is more appropriate for comparative purposes (Chung, 1997).

Thus, studying spectral properties of the Laplacian matrix offers an alternative way of getting insight into organization of the corresponding network (Merris, 1994). This implies solving the equation
\begin{equation}
{\bf {\hat L}} {\bf x}_i = \lambda_i {\bf x}_i ,
\end{equation}  
which determines the eigenvectors ${\bf x}_i$ and the corresponding eigenvalues $\lambda_i$. Since ${\bf {\hat L}}$ can be expressed as a product 
\begin{equation}
{\bf {\hat L}} = {\bf B}{\bf B}^T,
\label{BB} 
\end{equation}
where $\bf B$ is the incidence matrix, whose rows are indexed by the vertices and whose columns are indexed by the edges of the network, all the eigenvalues of the normalized Laplacian are real and non-negative. Furthermore, by construction the sum of entries in all the rows (or columns) of ${\bf {\hat L}}$ is zero which reduces dimensionality of this matrix (making it singular) and, thus, results in one zero eigenvalue representing the most collective mode (Dro\.zd\.z, Kwapie\'n, Speth \& W\'ojcik, 2002). A null hypothesis of purely random links in such networks thus corresponds to the Wishart ensemble of random matrices with the reduced rank (Janik \& Nowak, 2003).

In Fig.~7 the eigenvalue distributions for the normalized Laplacian matrices constructed from the strength $s$ for all the nine networks considered above and for the additional one composed of 2220 nodes including all papers by HES and also all papers by each of the six authors (MA, AB, SB, SH, CT, TV), whose own networks overlap with the one of HES and in separation are shown in Figs.~4-6. This extended network thus already involves many Stanley $\#2$ nodes and is denoted as $\it All (Stanley)$.  Consistent with the structure of the normalized Laplacian matrix, Eq.(~\ref{nlaplace}), the eigenvalues are centered on both sides of unity and one zero eigenvalue is of course always present. Some correlation between the relative locations of these eigenvalues and the weighted degree distributions of the corresponding networks in Figs.~1-6 can also be seen. When the central node constitutes an outlier in the degree distribution, the other eigenvalues of $\bf {\hat L}$, while developing a gap with respect to the zero mode, are spread more uniformly as compared to the case of no outlier. In the latter case, connectivity of the nodes in the corresponding network is larger, which drives a larger fraction of eigenvalues to be concentrated closer to unity as it is consistent with the Wishart-type product structure (Eq.(\ref{BB})) of the matrix ${\bf {\hat L}}$.

\begin{figure}[!h]
\centering
\includegraphics[scale=0.42]{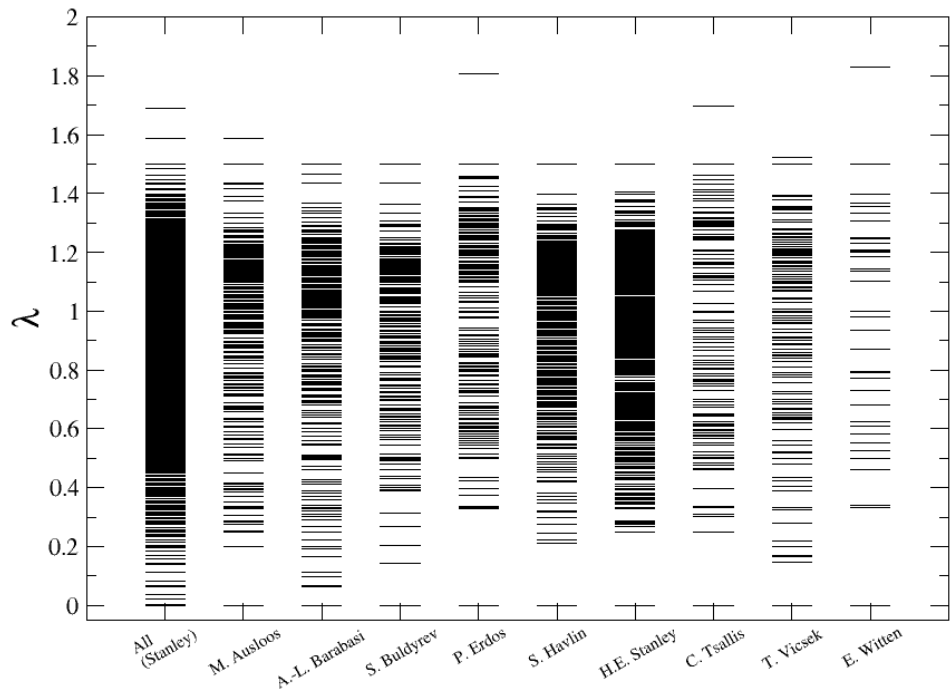}
\caption{Eigenvalue spectra of the normalized Laplacian matrices generated from the strength $s_{ij}$ of links as in Eq.(\ref{strength}) for the scientific collaboration networks determined by all the publications involving the authors listed along the horizontal axis. The additional {\it All (Stanley)} network results from all publications by Stanley and by other six authors (Ausloos, Barab\'asi, Buldyrev, Havlin, Tsallis, and Vicsek) whose own collaboration networks, overlapping with Stanley's, in separation are shown in Figs.~4-6.}
\label{fig7}
\end{figure}

A very valuable insight into organization of networks is offered by the eigenvectors $\bf x_i$ with components $\{ x^j_i \}$ ($\sum_j |x^j_i|^2 = 1$) as they reflect composition of the orthogonal 'modes' in the network (Kwapie\'n \& Dro\.zd\.z, 2012) and may thus project out communities within the corresponding network (Lancichinetti, Fortunato \& Kert\'esz, 2009). The potential of such a procedure for the most involved case considered here of {\it All (Stanley)} is demonstrated in Fig.~8, which shows the distributions of the eigenvector components $x^i_j$ for the six eigenvectors corresponding to the eigenvalues starting from the lowest ($i=1$) and going upwards to $i=6$. The $i=1$ eigenvector is seen to represent the most collective structure as essentially all its components contribute visibly. The arrows indicate to whom the largest contributions belong. In $i=1$ they are seen to belong to the dominant hubs and the largest contribution here is due to HES. The other eigenvectors are already much more selective (and involve the negative sign as well). $i=2$ is dominated by Ausloos, while $i=3$ by Tsallis and by their collaborators, respectively. $i=4$ again represents a mixture of the names from $i=1$, but here the Buldyrev's sector somewhat overtakes. $i=5$ and $i=6$ on the other hand project out the Barabasi and Vicsek sectors and their own collaborative relations. Such a composition of the consecutive eigenvectors is quite understandable when mutual overlaps and convolutions of the sub-communities inside the {\it All (Stanley))} network are explicitly inspected. Thus it indicates potential utility of the procedure sketched here for the community detection in complex networks of larger size, where a visual identification of such effects is the most likely prohibitively impractical.

\vspace{0.5cm}
\begin{figure}[!h]
\centering
\hspace{-1cm}
\includegraphics[scale=0.5]{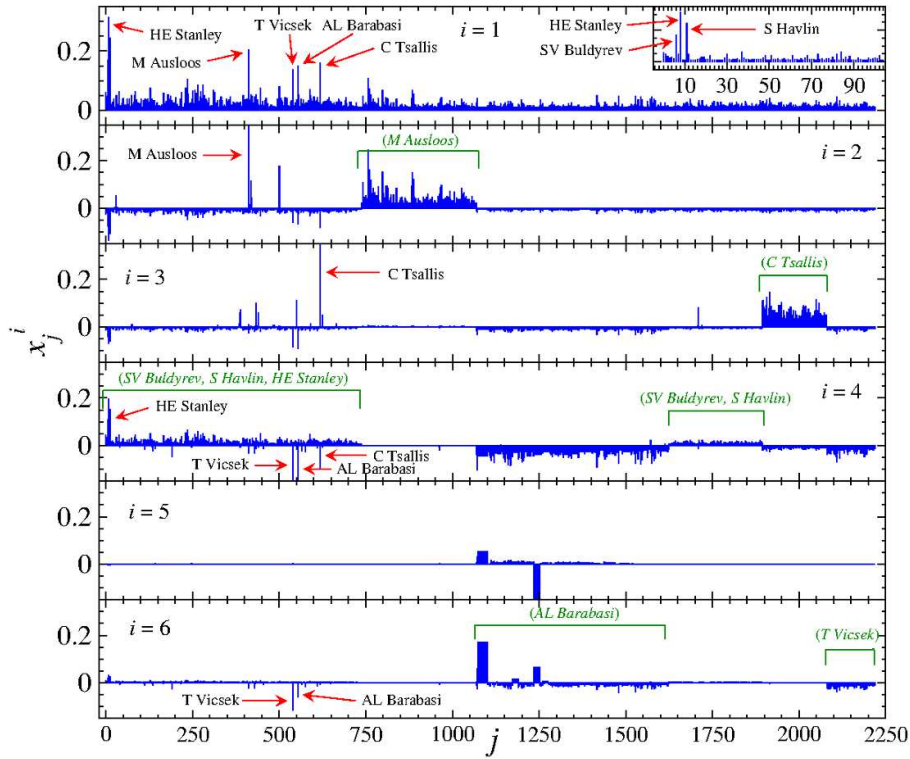}
\caption{Components $x^i_j$ of the six consecutive eigenvectors $i=1,...,6$,
starting with the 'zero mode' upwards, of the normalised Laplacian matrix $\bf {\hat L}$ defined by Eq.~(\ref{nlaplace}), whose entries are constructed from the collaborative strength $s_{ij}$ (Eq.~(\ref{strength})) for the scientific collaboration network denoted as {\it All (Stanley)}.}
\label{fig8}
\end{figure}

\section{Summary} 

As the above explicit inspection of the scientific collaboration networks by several renown scholars shows, their structure reveals a rich diversity of the quantifiable characteristics. This points to the utility of such a representation for the scientometric and bibliometric studies, to quantify characteristics of communities, and offers an interesting perspective to inspect mechanisms that stay behind the development and the spreading tendencies of the contemporary scientific collaboration. 

From the complex networks point of view, the most interesting network is the one of H. Eugene Stanley (Stanley $\#1$), which is of a large size and develops a visibly hierarchical organization through all its scales. However, this is seen in the weighted network representation and confirmed by the resulting degree and by the clustering coefficient distributions. A particularly significant fact in this context is that the central node defining this network, HES, obeys the same functional form in these two distributions as all the other nodes belonging to this network. Interestingly, as Fig.~3 shows, this network attained such an organization already in around 1995 and largely preserved through the next 20 years even though it more than doubled in size until present. Similar characteristics are observed in the Ausloos $\#1$ and, to some extent, also in the Havlin $\#1$ networks, though these two are about a factor of two smaller and the latter in addition strongly overlaps with the HES network. 

Self-organizing growth of such networks demands of course a special balance between the number of incoming new nodes in time (with new publications) and appearance of the new links (co-authorships) among the nodes already belonging to this network, with some elements of the preferential attachment in distributing the links so that the hierarchical organization builds up and the weighted degree distribution assumes the scale-free form. An existing relevant model, in which the usual preferential attachment (Barab\'asi \& Albert, 1999) is extended to the rule "busy get busier" (Barthelemy, Barrat, Pastor-Satorras \& Vespignani, 2005), where the new nodes connect more likely to the nodes carrying larger weights and which are more central in terms of the strength of interactions, offers some quantitative insight into the underlying mechanism. In particular, it indicates that, in networks with such characteristics as the one of HES, the spread of weight gets stimulation to multiplicative bursts over the neighbouring nodes and this leads to a balanced distribution of interconnections among them such that the scale-free Stanley $\#1$ hierarchy develops. 

The unquestionably interdisciplinary and diversified character of the corresponding HES (and largely also Ausloos) scientific activity may constitute a significant factor that facilitates and is even likely to favour a spontaneous generation of such interconnections as it is needed for their balanced growth. In the cases of more specialized activity, the growth of the corresponding networks may progress through a smaller number of the mutual interconnections. In the latter case, this is thus expected to result in a deficit of such interconnections relative to the number of links acquired by the central hub and, therefore, such a hub is more likely to maintain an outlier position in the degree distribution. Relating Stanley $\#1$ and Ausloos $\#1$ networks to those of Erd\H{o}s or Witten, whose scientific activities are more specialized, indicates that this may be a relevant element that boosts a potential in this kind of networks to become hierarchical, indeed.

Of course, development of such a hierarchy involves and is embedded in many overlapping communities (Palla, Der\'enyi, Farkas \& Vicsek, 2005). A sample spectral analysis in Section~6 of the normalised Laplacian matrix corresponding to the extended scientific collaboration network, including also a number of Stanley $\#2$ nodes, points to the usefulness of such a procedure for identifying and for characterising the related component communities in the weighted networks.

\end{document}